\documentclass{INTERSPEECH2023}
\usepackage{amsmath,graphicx}
\usepackage{amsmath,amsfonts}
\usepackage{algorithmic}
\usepackage{algorithm}
\usepackage{array}
\usepackage{textcomp}
\usepackage{subfig}
\usepackage{stfloats}
\usepackage{url}
\usepackage{verbatim}
\usepackage{tikz}
\usepackage{cite}
\usetikzlibrary{positioning,shapes}

\usepackage{mathtools}
\newcommand{\norm}[1]{\left\lVert#1\right\rVert}
\usepackage{xfrac}

\newcommand{\D}[1]{\mathrm{d}{#1}}
\newcommand{\submin}[0]{_\text{min}}
\newcommand{\submax}[0]{_\text{max}}

\DeclareMathOperator*{\argmin}{arg\,min}

\usepackage[font=small,skip=0pt]{caption}
\usepackage[nolist, nohyperlinks]{acronym}
\begin{acronym}
\acro{bb}[BB]{Brownian bridge}
\acro{sgm}[SGM]{score-based generative model}
\acro{snr}[SNR]{signal-to-noise ratio}
\acro{gan}[GAN]{generative adversarial network}
\acro{vae}[VAE]{variational autoencoder}
\acro{ddpm}[DDPM]{denoising diffusion probabilistic model}
\acro{stft}[STFT]{short-time Fourier transform}
\acro{istft}[iSTFT]{inverse short-time Fourier transform}
\acro{sde}[SDE]{stochastic differential equation}
\acro{ode}[ODE]{ordinary differential equation}
\acro{ou}[OU]{Ornstein-Uhlenbeck}
\acro{ve}[VE]{Variance Exploding}
\acro{dnn}[DNN]{deep neural network}
\acro{pesq}[PESQ]{Perceptual Evaluation of Speech Quality}
\acro{se}[SE]{speech enhancement}
\acro{tf}[T-F]{time-frequency}
\acro{elbo}[ELBO]{evidence lower bound}
\acro{WPE}{weighted prediction error}
\acro{PSD}{power spectral density}
\acro{RIR}{room impulse response}
\acro{SNR}{signal-to-noise ratio}
\acro{LSTM}{long short-term memory}
\acro{POLQA}{Perceptual Objectve Listening Quality Analysis}
\acro{SDR}{signal-to-distortion ratio}
\acro{ESTOI}{Extended Short-Term Objective Intelligibility}
\acro{ELR}{early-to-late reverberation ratio}
\acro{TCN}{temporal convolutional network}
\acro{DRR}{direct-to-reverberant ratio}
\acro{nfe}[NFE]{number of function evaluations}
\acro{rtf}[RTF]{real-time factor}
\end{acronym}

\def\BibTeX{{\rm B\kern-.05em{\sc i\kern-.025em b}\kern-.08em
    T\kern-.1667em\lower.7ex\hbox{E}\kern-.125emX}}

\title{Reducing the Prior Mismatch of Stochastic Differential Equations for Diffusion-based Speech Enhancement}
\name{Bunlong Lay$^1$, Simon Welker$^{1,2}$, Julius Richter$^1$, Timo Gerkmann$^1$\thanks{\scriptsize We acknowledge the support by DASHH (Data Science in Hamburg - HELMHOLTZ Graduate School for the Structure of Matter) with the Grant-No. HIDSS-0002 and the German Research Foundation (DFG) in the transregio project Crossmodal Learning (TRR 169).}}
\address{
  $^1$Signal Processing (SP), University of Hamburg, Germany \\
  $^2$Center for Free-Electron Laser Science, DESY, Hamburg, Germany}
\email{\{bunlong.lay, simon.welker, julius.richter, timo.gerkmann\}@uni-hamburg.de}

\begin{document}
\maketitle
\begin{abstract}
Recently, score-based generative models have been successfully employed for the task of speech enhancement. A stochastic differential equation is used to model the iterative forward process, where at each step environmental noise and white Gaussian noise are added to the clean speech signal. While in limit the mean of the forward process ends at the noisy mixture, in practice it stops earlier and thus only at an approximation of the noisy mixture. This results in a discrepancy between the terminating distribution of the forward process and the prior used for solving the reverse process at inference. In this paper, we address this discrepancy and propose a forward process based on a Brownian bridge. We show that such a process leads to a reduction of the mismatch compared to previous diffusion processes. More importantly, we show that our approach improves in objective metrics over the baseline process with only half of the iteration steps and having one hyperparameter less to tune.
\end{abstract}

\section{Introduction}
\label{sec:intro}

Speech enhancement aims to recover the clean speech signal from a noisy mixture that is corrupted by environmental noise~\cite{hendriks2013dft}. Classical approaches try to exploit statistical relations of the clean speech signal and the environmental noise~\cite{gerkmann2018book_chapter}. Numerous machine learning methods have been proposed that treat speech enhancement as a discriminative learning task \cite{wang2018supervised, luo2019conv}.

Different from discriminative approaches that learn a direct mapping from noisy to clean speech, generative approaches learn a prior distribution over clean speech data. 
Recently, so-called \emph{score-based generative models} (or \emph{diffusion models}) were introduced to the task of speech enhancement \cite{lu2021study, lu2022conditional, paper, journal, serra2022universal}. The idea is to iteratively add Gaussian noise to the data using a discrete and fixed Markov chain called \emph{forward process}, thereby transforming data into a tractable distribution such as a normal distribution. Then, a neural network is trained to invert this diffusion process in a so-called \emph{reverse process} \cite{ho2020denoising}. When the step size between two discrete Markov chain states is taken to zero, the discrete Markov chain becomes a continuous-time \ac{sde} under mild constraints. Utilizing SDEs offers more flexibility and opportunities than approaches based on discrete Markov chains \cite{song2021sde}. For example, SDEs allow to use general-purpose \ac{sde} solvers to numerically integrate the reverse process, impacting the performance and number of iteration steps. An SDE can be interpreted as a transformation between two given distributions, where one is called the initial distribution and the other the terminating distribution. In the case of speech enhancement, we transform between the distribution of clean speech data and the distribution of noisy mixture data. Under mild constraints, we can find for each forward SDE a reverse SDE inverting the forward SDE \cite{anderson1982reverse, haussmann1986time}. This reverse SDE starts from a noisy mixture and ends at the clean speech. It can be therefore used for speech enhancement. 

Currently, for the task of speech enhancement, there are different approaches that integrate the corruption of environmental noise in the diffusion process \cite{lu2022conditional, paper, journal}. To compensate for non-Gaussian noise characteristics, these approaches use an interpolation between clean speech and noisy speech data along the forward process. In \cite{paper, journal} a continuous-time SDE is used, which includes a drift term that allows the transformation between clean and noisy speech.
Interestingly, the mean of the process in \cite{paper, journal} evolves from clean speech perfectly to noisy speech only for an infinitely long forward diffusion process.
In practice, however, the mean of the forward process ends at an approximation of the noisy speech data. Therefore, when solving the reverse SDE to perform speech enhancement, there exists a mismatch between the terminating distribution of the forward process and the initial distribution of the reverse process \cite{journal}.  We call the initial distribution of the reverse process the prior distribution of the generative model and the corresponding mismatch the \emph{prior mismatch}. Moreover, the SDEs in \cite{paper, journal, popov2022diffusionbased, popovGradTTSDiffusionProbabilistic2021} includes a stiffness parameter controlling the pull of the terminating distribution of the forward process and the prior distribution. Consequently, this stiffness parameter determines the degree of the resulting prior mismatch. Increasing the stiffness reduces the prior mismatch, but may also negatively affect the speech enhancement performance as the reverse process may become unstable \cite[Section II D]{journal}.

To overcome this limitation, we seek to reduce the prior mismatch without destabilizing the reverse process. To this end, we propose to replace the forward process in \cite{paper, journal} with an SDE based on a Brownian bridge process. A Brownian bridge seems suitable for this purpose because it has fixed starting and end points and follows a Brownian motion in between. We show that the resulting diffusion process does not only drastically decrease the prior mismatch, but also eliminates the dataset-dependent and hard-to-tune stiffness parameter of the SDE in \cite{paper, journal}. In the experiments, we demonstrate that using the proposed SDE outperforms the baseline SDE while having one hyperparameter less to tune and using only half as many function evaluations \footnote{code online available \url{https://github.com/sp-uhh/sgmse-bbed}}.

\section{Background}
The task of speech enhancement is to estimate the clean speech signal $\mathbf S$ from a noisy mixture $\mathbf Y = \mathbf S+\mathbf N$, where $\mathbf N$ is environmental noise. All variables in bold are the coefficients of a complex valued \ac{stft}, e.g. $\mathbf Y \in \mathbb{C}^d$ and $d=KF$ with $K$ number of \ac{stft} frames and $F$ number of frequency bins.

\subsection{Stochastic Differential Equations} \label{sec:sde}
Following the approach in \cite{paper, journal}, we model the forward process of the score-based generative model with an \ac{sde} defined on $0 \leq t < T_{\text{max}}$:
\begin{equation} \label{eq:fsde}
    \D{\mathbf X_t} =
       \mathbf f(\mathbf X_t, \mathbf Y) \D{t}
        + g(t)\D{{\mathbf w}},
\end{equation}
where $\mathbf w$ is the standard Wiener process \cite{kara_and_shreve}, $\mathbf X_t$ is the current process state with initial condition $\mathbf X_0 = \mathbf S$, and $t$ a continuous diffusion time-step variable describing the progress of the process ending at the last diffusion time-step $T_{\text{max}}$.
Moreover, $\mathbf f(\mathbf X_t, \mathbf Y) \D{t}$ can be integrated by Lebesgue integration \cite{rudin}, and $g(t)\D{{\mathbf w}}$ follows Ito integration \cite{kara_and_shreve}. 
The functions $\mathbf f(\mathbf X_t, \mathbf Y)$ and $g(t)$ are called drift and diffusion coefficient, respectively. The diffusion coefficient $g$ regulates the amount of Gaussian noise that is added to the process, and the drift $\mathbf f$  affects mainly in the case of linear SDEs the mean of $\mathbf X_t$ (see \cite[(6.10)]{kara_and_shreve}). The process state $\mathbf X_t$ follows a Gaussian distribution \cite[Ch. 5]{sarkkaAppliedStochasticDifferential2019}, called the \emph{perturbation kernel}:
\begin{equation}
\label{eq:perturbation-kernel}
    p_{0t}(\mathbf X_t|\mathbf X_0, \mathbf Y) = \mathcal{N}_\mathbb{C}\left(\mathbf X_t; \boldsymbol \mu(\mathbf X_0, \mathbf Y, t), \sigma(t)^2 \mathbf{I}\right).
\end{equation}
By Anderson \cite{anderson1982reverse}, each forward SDE as in \eqref{eq:fsde} can be associated to a reverse SDE:
\begin{equation}\label{eq:plug-in-reverse-sde}
    \D{\mathbf X_t} =
        \left[
            -\mathbf f(\mathbf X_t, \mathbf Y) + g(t)^2\mathbf  \nabla_{\mathbf X_t} \log p_t(\mathbf X_t|\mathbf Y)
        \right] \D{t}
        + g(t)\D{\bar{\mathbf w}}\,,
\end{equation}
where
$\D{\bar{\mathbf w}}$ is a Wiener process going backwards in time. In particular, the reverse process starts at $t=T$ and ends at $t=0$. Here $T < T_{\text{max}}$ is a parameter that needs to be set for practical reasons as the last diffusion time-step $T_{\text{max}}$ is only reached in limit.
The \emph{score function} $\nabla_{\mathbf X_t} \log p_t(\mathbf X_t|\mathbf Y)$ is approximated by a neural network called \emph{score model} $s_\theta(\mathbf X_t, \mathbf Y, t)$, which is parameterized by a set of parameters $\theta$.
Assuming that $s_\theta$ is available, we can generate an estimate of the clean speech $\mathbf X_0$ from $\mathbf Y$ by solving the reverse SDE.

The prior mismatch discussed in this paper is defined by the difference of $\boldsymbol \mu(\mathbf X_0, \mathbf Y, T)$ to $\mathbf Y$.
In this work and previous work \cite{paper, journal}, we consider only SDEs where the mean is of the form 
\begin{equation} \label{eq:interp}
\boldsymbol \mu(\mathbf X_0, \mathbf Y, t) = (1-k(t)) \mathbf X_0 + k(t) \mathbf Y\,,\end{equation}
where $0 \leq k(t) < 1$ is an increasing function. In the sequel, we will simply write $\boldsymbol \mu(t)$ for brevity. The mismatch of such an SDE is determined by $k(T)$ and we call $k(T)$ the \emph{maximal interpolation factor} (MIF) for the rest of the paper. It is desired that the MIF $k(T)$ is close to 1 and we will see in the following sections to which degree this goal is met.

\section{Design choices of different SDEs}
\subsection{Ornstein-Uhlenbeck with Variance Exploding (OUVE)} \label{sec:ouve} 
In \cite{paper, journal} an SDE is used with the drift coefficient $f(\mathbf X_t, \mathbf Y)$ and diffusion coefficient $g(t)$ defined as
\begin{align} \label{eq:ouve-sde}
    \mathbf f(\mathbf X_t, \mathbf Y) &= \gamma(\mathbf Y-\mathbf X_t), \\ \label{eq:ouve-diffusion}
    g(t) &= \sigma\submin  \left(\frac{\sigma\submax}{\sigma\submin}\right)^t \sqrt{2\log\left(\frac{\sigma\submax}{\sigma\submin}\right)}~, 
\end{align}
for $0 \leq t\leq T<T_\text{max} = \infty$ and parameters $\gamma, \sigma\submin$, $\sigma\submax \in \mathbb R_+$. Such a drift term is typical for an Ornstein-Uhlenbeck process \cite{kara_and_shreve}, whereas the diffusion coefficient is taken from the so-called \emph{Variance Exploding} SDE \cite{song2021sde}. Thus, we call the baseline SDE \emph{Ornstein-Uhlenbeck with Variance Exploding} (OUVE). A reparameterization of \eqref{eq:ouve-diffusion} with $\sigma \submax \coloneqq k \sigma \submin$ and $c \coloneqq \sigma^2 \submin 2\log(\frac{\sigma \submax}{\sigma \submin})$ yields
\begin{equation} \label{eq:ouve-diffusion-repara}
    g(t) = \sqrt{c}k^t, ~~\text{where } c,k > 0.
\end{equation}
We argue that this equivalent representation of the diffusion coefficient may increase the intuition of \eqref{eq:ouve-diffusion}, as $\sqrt{c}$ simply scales the diffusion coefficient and $k$ is the base of the exponential term. We will simply use the parameterization of Eq. \eqref{eq:ouve-diffusion-repara} for the rest of this work.

The closed-form solution for the mean and variance of the perturbation kernel of this SDE are given by:

\begin{equation}
    \label{eq:std_ouve}
    \sigma(t)^2 = \frac{
        c\left(k^{2t} - \mathrm e^{-2\gamma t}\right)}
        {2(\gamma+\log(k))}
    \,,
\end{equation}

and 
\begin{equation}
\label{eq:mean_ouve}
   \boldsymbol \mu(t) = \mathrm e^{-\gamma t} \mathbf X_0 + (1-\mathrm e^{-\gamma t}) \mathbf Y
    \,.
\end{equation}
We see from \eqref{eq:mean_ouve} that for large $t\to \infty$, we have that $\mathbf X_t$ has mean $\mathbf Y$. However, as in practice we need to decide for a finite final diffusion time-step $T$, a certain difference between the mean of $\mathbf X_T$ and $\mathbf Y$ remains. If we parameterize the OUVE SDE as in \cite{journal}, i.e. $\gamma=1.5$ and $T=1$, then we find that the MIF is $k(T) = (1-e^{-1.5}) = 0.78$. As it is desired to have a MIF close to 1, we argue that the difference between $\boldsymbol \mu(T)$ and $\mathbf Y$ is relatively large.
Note that increasing $T$ for fixed $\gamma$ to obtain a better MIF is equivalent to fixing $T$ and increasing $\gamma$. Moreover, we have that increasing $\gamma$, yields a better MIF, but also worsens the performance of this approach, as the sampling from the reserve SDE becomes unstable \cite[Section II D]{journal}. 
Therefore, increasing the MIF $k(T) = 1-\mathrm e^{-\gamma T}$ for this SDE is not straightforward. 

\subsection{Brownian Bridge with Exponential Diffusion Coefficient (BBED)} \label{sec:bb}
In order to reduce the prior mismatch, we propose to employ an SDE that has a linear interpolation factor $k(t)=t$, where $ 0\leq t \leq T < T_{\text{max}}=1$.
Substituting $k(t)=t$ the mean of the SDE in \eqref{eq:interp} becomes
\begin{equation}
\label{eq:mean_bb}
    \boldsymbol \mu(t) = \left(1- t\right)\mathbf X_0 + t \mathbf Y
    \,.
\end{equation}
One can find an SDE with the following drift coefficient that has the desired mean from \eqref{eq:mean_bb} by solving \cite[(6.12)]{kara_and_shreve}
\begin{align} \label{eq:bb-drift}
    \mathbf f(\mathbf X_t, \mathbf Y) &= \frac{\mathbf Y-\mathbf X_t}{1-t}.
\end{align}
 Comparing \eqref{eq:mean_bb} and \eqref{eq:interp}, we see that the MIF is $k(T)=T$. Note, that the choice of $T<1$ is limited due to numerical stability as we divide by $(1-t)$ in \eqref{eq:bb-drift}. However, it is still possible to achieve a much better MIF compared to the MIF of the OUVE SDE, as we will see in Section \ref{sec:res:reduce_mm}. 

For a fair comparison to the OUVE SDE, we want to utilize the same diffusion coefficient as from the OUVE SDE in Eq. \eqref{eq:ouve-diffusion-repara}.
The resulting variance can be computed from \cite[(6.11)]{kara_and_shreve}:
\begin{align} \label{eq:bbed:var}
   \hspace{-0.65em} \sigma(t)^2 &=  (1-t)c\left[(k^{2t}-1+t) + \log(k^{2k^2})(1-t)E \right], \\
    E &= \text{Ei}\left[2(t-1)\log(k)\right] - \text{Ei}\left[-2\log(k)\right],
\end{align}
where $\text{Ei}[\cdot]$ denotes the exponential integral function \cite{bender78:AMM}. The variance trajectory exhibits one peak and vanishes for $t=0$ and $t=1$. The position of the peak is solely determined by $k$, where larger $k$ shifts the peak closer to $t=1$.

In the literature, SDEs that linearly transform the starting condition ($\mathbf X_0 = \mathbf S$ and zero variance for $t=0$) to the terminal condition ($\mathbf X_T = \mathbf Y$ and zero variance for $t=1$) with a constant diffusion coefficient of $g(t)=1$ are called Brownian bridges \cite{kara_and_shreve}. As the SDE with drift coefficient \eqref{eq:bb-drift} and diffusion coefficient \eqref{eq:ouve-diffusion-repara} differs from that definition only in the diffusion coefficient, we call the SDE a \emph{Brownian Bridge with Exponential Diffusion coefficient} (BBED).

\begin{figure}[t]
\centering
\raggedright
\includegraphics[trim=0.2cm 0 0 0, scale=0.9]{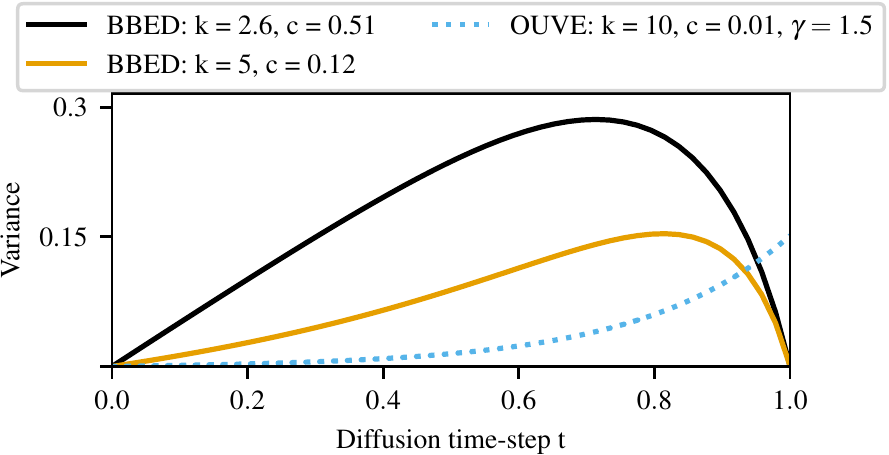}
\vspace{0.5em}
\caption{Variance evolution of BBED and OUVE. Solid curves are variances for BBED with different parameterizations. The dotted curve is the variance of the OUVE with parameterization as in Section \ref{sec:exp:sde}.}
\label{fig:variance}
\end{figure}

\begin{figure}[t]
\centering
\raggedright
\includegraphics[trim=0.2cm 0 0 0, scale=0.9]
{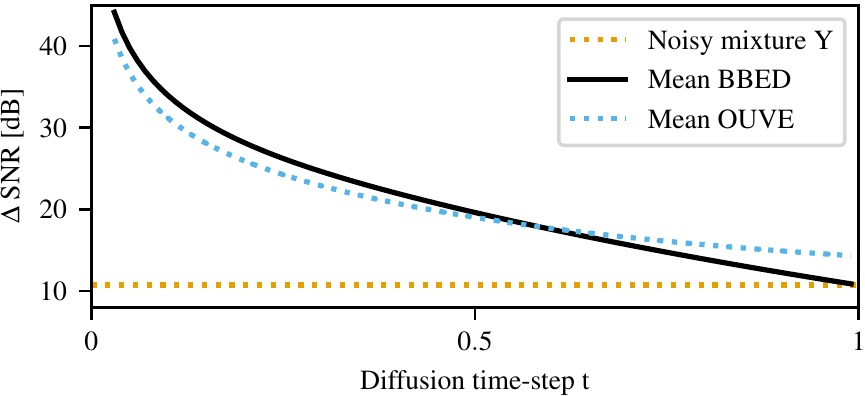}
\vspace{0.5em}
\caption{Black and blue curves are the averaged $\Delta \text{SNR}(\mathbf \mu(t))$ as defined in \eqref{eq:dsnr} of the mean evolutions of BBED and OUVE for the WSJ0-CHiME3 test set. The dotted yellow line is the SNR of $\mathbf Y$. The OUVE SDE is parameterized with $\gamma=1.5$ as described in Section \ref{sec:exp:sde}.}
\label{fig:mean}
\end{figure}    

\section{Experimental setup} \label{sec:exp}
To allow a fair comparison between BBED SDE with OUVE SDE, we train the corresponding score models with the same configuration and follow the experimental setup from \cite[Section V]{journal}.

\subsection{Training}
For the score model $s_\theta(\mathbf X_t, \mathbf Y, t)$,  we employ the Noise Conditional Score Network (NCSN++) architecture (see \cite{journal, song2021sde} for more details). The network is optimized based on denoising score matching:
\begin{equation}\label{eq:training-loss}
      \argmin_\theta \mathbb{E}_{t,(\mathbf X_0,\mathbf Y), \mathbf Z, \mathbf X_t|(\mathbf X_0,\mathbf Y)} \left[
        \norm{\mathbf s_\theta(\mathbf X_t, \mathbf Y, t) + \frac{\mathbf Z}{\sigma(t)}}_2^2
    \right],
\end{equation}
where $\mathbf X_t =  \boldsymbol \mu(t) + \sigma(t) \mathbf Z$ with $\mathbf Z \sim \mathcal N_{\mathbb{C}}(\mathbf 0, \mathbf I)$. We train the network with the ADAM optimizer \cite{kingma2015adam} with a learning rate of $10^{-4}$ and a batch size of 16. An exponential moving average of network parameters is tracked with a decay of 0.999, to be used for sampling \cite{paper,song2021sde}. We train for 250 epochs and log the averaged PESQ value of 10 random files from the validation set during training and select the best-performing model for evaluation. Experiments are conducted on an NVIDIA A6000 and training lasts for approximately 4 days.

\subsection{Dataset and input representation} \label{sec:exp:data}

We use the same WSJ0-CHiME3 dataset as in \cite{journal}. This dataset mixes clean speech utterances from the Wall Street Journal (WSJ0) dataset \cite{datasetWSJ0} to noise signals from the CHiME3 dataset \cite{barker2015third} with an uniformly sampled \ac{snr} between 0 and 20$\,$dB. The dataset is split into a train (12777 files), validation (1206 files) and test set (615 files).

Each file from the WSJ0-CHiME3 dataset is converted into a complex STFT representation with a window size of 510, resulting in 256 frequency bins, a hop size of 128 and a periodic Hann window. We randomly crop the \ac{stft} representation to a length of 256 frames at each training step. To compensate for the typically heavy-tailed distribution of \ac{stft} speech magnitudes~\cite{gerkmann2010empirical}, as in \cite{journal}, each complex coefficient $c$ of the \ac{stft} representation is transformed via $\beta |c|^\alpha \mathrm e^{i \angle(c)}$ with $\beta=0.15$ and $\alpha=0.5$.

\subsection{Sampling and metrics} \label{sec:exp:sampling}
For the baseline OUVE SDE and the proposed BBED SDE we use the same sampler settings for a fair comparison. We use a Predictor-Corrector scheme as in \cite{journal, song2021sde}, where the Predictor is the Euler-Maruyama method \cite{sarkkaAppliedStochasticDifferential2019} and the Corrector is the Annealed Langevin Dynamics (ALD) method \cite{song2021sde}. As in \cite{journal}, the step size for ALD is chosen as $0.5$ and the number of reverse steps is $30$. Equivalently, the step size in the reverse process is $h=T/30$, where $T$ is set for the OUVE SDE and BBED SDE individually (see Section \ref{sec:exp:sde}). For the reverse process, we set the reverse starting time at $t_{\text{rs}}=T$. We also report results when experimenting with the reverse starting times $t_{\text{rs}} < T$ in Section \ref{sec:ablation} while keeping the step size $h=T/30$ fixed.

We evaluate the performance on perceptual metrics, wideband PESQ \cite{rixPerceptualEvaluationSpeech2001} and POLQA \cite{polqa2018}, on energy-based metrics SI-SDR, SI-SIR and SI-SAR \cite{lerouxSDRHalfbakedWell2018} and on intelligibility metric ESTOI \cite{jensen2016algorithm}.

\subsection{OUVE and BBED} \label{sec:exp:sde} 
As in \cite{journal}, the parameters $T$, $\sigma \submin$, $\sigma \submax$ and $\theta$ were already tuned by a grid search. Therefore, we set as in \cite{journal} $T=1$ and $\gamma=1.5$, and the diffusion coefficient parameters in Eq. \ref{eq:ouve-diffusion} are set to  $\sigma \submin = 0.05$ and $\sigma \submax = 0.5$, or in the equivalent representation in Eq. \ref{eq:ouve-diffusion-repara}, we set $k=10$ and $c=0.01$.

For the BBED SDE, we search for the largest $T$ in $\{0.9, 0.99, 0.999, 0.9999 \}$ so that training and inference is numerically stable. The parameter $k$ in \eqref{eq:ouve-diffusion-repara} is determined as the empirically optimal choice of $K \coloneqq \{0.02, 0.2, 0.6, 1.1, 1.5, 2.6, 5, 27 \}$. The values of the grid have been chosen in such a way that the resulting variances have their peaks ranging from $0.2$ to $0.9$. For example, the resulting variance for $k=5$ has its maximum at $0.8$, the variance for $k=2.6$ has its maximum at $0.7$, etc. For each $k\in K$, we set the normalization factor $c$ so that the variances admit a maximum value of either 0.15 or 0.3. This choice is based on the OUVE SDE parameterization also having a maximum value of $0.15$. Exemplary, we plot two parameterizations of the variance of the BBED SDE in Fig. \ref{fig:variance}.

\begin{table*}[t!]
    \vspace*{-0.0cm}
\centering
\caption{Speech enhancement results (average and standard deviation over the test set) obtained for WSJ0-CHiME3. The OUVE SDE is parameterized as described in \ref{sec:exp:sde} and the BBED SDE is parameterized with $k=2.6, c=0.51$. $t_{\text{rs}}$ denotes the reverse starting time as defined in Section \ref{sec:exp:sampling}.}
\vspace{0.5em}
\begin{tabular}{c|ccccccc}
SDE & POLQA & PESQ & ESTOI & SI-SDR [dB] & SI-SIR [dB] & SI-SAR [dB]  \\
\hline
\hline
Mixture & $2.63 \pm 0.67$ & $1.70 \pm 0.49$ & $0.78 \pm 0.14$ & $10.0 \pm 5.7$ & $10.0 \pm 5.7$ & - \\
\hline
Baseline OUVE \cite{journal} & $3.71 \pm 0.51$ & $2.92 \pm 0.53$ & $0.92 \pm 0.05$ & $17.78 \pm 4.57$ & $31.51 \pm 4.9$ & $18.00 \pm 4.65$ \\
\hline
BBED $t_{\text{rs}}=0.5$ & $ 3.97  \pm 0.48 $ &   $3.05 \pm 0.53$ & $0.93 \pm 0.05$& $18.96 \pm 4.28$ & $31.42 \pm 5.19$  & $19.28 \pm 4.36$ \\
\hline
BBED $t_{\text{rs}}=0.999$ & $ \mathbf{4.01  \pm 0.49} $ &   $\mathbf{3.08 \pm 0.57}$ & $\mathbf{0.94 \pm 0.05}$& $\mathbf{19.26 \pm 4.43}$ & $\mathbf{31.64 \pm 5.08}$  & $\mathbf{19.59 \pm 4.53}$ \\
\hline
\hline
\end{tabular}
    \label{tab:results:wsj0}
\end{table*}

\section{Results} \label{sec:results}
First, we present the results when parameterizing the BBED SDE as described in Section \ref{sec:exp:sde}. Second, we discuss if the proposed BBED SDE reduces the prior mismatch compared to the baseline OUVE SDE.
Last, we discuss the performance differences in terms of objective metrics, number of iterations in the reverse process and subjective differences of the OUVE SDE and BBED SDE.

\subsection{Parameterization of the BBED SDE} \label{sec:res:para}
When training and testing the score-model with the BBED SDE with different $k \in K$, 
we argue that it is beneficial to have the variance maximum towards the end of the forward process, as the Gaussian noise would better mask the speech features corrupted by the environmental noise. At the same, if the variance maximum is too close to the end of the forward process, which is at $t=1$, then the diffusion coefficient becomes numerically large and consequently the reverse process may become unstable. Empirically, we found that $k=2.6$ with maximum variance $0.3$ results in the best performance (see Fig. \ref{fig:variance} black line). When training and testing the score-model with the BBED SDE with different $T \in \{0.9, 0.99, 0.999, 0.9999\}$, we found that $T=0.999$ is the largest value that causes no numerical issues.

\subsection{Reducing the prior mismatch} \label{sec:res:reduce_mm}
As we set for the BBED SDE $T=0.999$, we have that the MIF is $k(T)=0.999$. This is much closer to $1$ than the MIF of $0.78$ achieved by the OUVE SDE as discussed in Section \ref{sec:ouve}. We illustrate this prior mismatch in terms for \ac{snr} in Fig. \ref{fig:mean}. To this end, let $y'$ and $s$ be time-domain signals, where $s$ is the clean speech signal and $y'$ is any clean speech signal corrupted with environmental noise. We define the SNR($\mathbf Y'$, $\mathbf S$) to be $20\log_{10}\frac{||s||_2}{||y'-s||_2}$, $|| \cdot ||_2$ denotes the  $\ell ^2$ norm. In Fig. \ref{fig:mean}, we averaged
\begin{equation} \label{eq:dsnr}
    \Delta \text{SNR}(\mathbf \mu(t)) \coloneqq \text{SNR}(\mathbf \mu(t), \mathbf S) - \text{SNR}(\mathbf Y, \mathbf S)
\end{equation}
for the BBED SDE and OUVE SDE over the WSJ0-CHiME3 test set.
We have that $\text{SNR}(\mathbf \mu(t), \mathbf S)$ approaches $ \text{SNR}(\mathbf Y, \mathbf S)$ if $\mathbf \mu(t) = \mathbf Y$. This is the case for the BBED SDE as it can be observed in Fig. \ref{fig:mean}. In comparison, we find that the OUVE SDE differs to the $\text{SNR}(\mathbf Y, \mathbf S)$ by $3.6$ dB at $t=1$ in Fig. \ref{fig:mean}, showing that the BBED SDE indeed reduces the prior mismatch compared to the OUVE SDE.

\subsection{OUVE vs. BBED} \label{sec:ablation}
In Tab. \ref{tab:results:wsj0} we show that BBED outperforms OUVE in all reported metrics. When listening to enhanced files generated by BBED and OUVE, we observe that the enhanced files generated by BBED contain less background noise and breathing artifacts than the enhanced files generated by OUVE. We provide some listening examples in the supplementary material\footnote{\url{https://www.inf.uni-hamburg.de/en/inst/ab/sp/publications/sgmse-bbed}}. %

Remarkably, when experimenting with $t_\text{rs}$ we found that the BBED SDE largely maintains performance when changing $t_\text{rs} = T = 0.999$ to $t_\text{rs} = 0.5$ as it can be seen in Tab. \ref{tab:results:wsj0}. This is in contrast to OUVE SDE which loses $0.31$ in PESQ when we set $t_\text{rs}=0.5$. %
Since we keep the reverse step size $h=T/30$ fixed when starting inference at $t_\text{rs}=0.5$, the number of iterations is halved with only negligible performance loss for the BBED SDE as compared to when starting the reverse process at $t_{\text{rs}}=0.999.$ In particular BBED even outperforms OUVE when only using half as many iterations for enhancement.

The proposed BBED SDE has a different drift coefficient compared to the OUVE SD (compare Eq. \eqref{eq:bb-drift} and Eq. \eqref{eq:ouve-sde}), which results in different mean evolutions (see Fig. \ref{fig:mean}) and different variance evolutions (see Fig. \ref{fig:variance}). Thus, there could be various reasons why the BBED SDE outperforms the OUVE SDE. 
We hypothesize that a much higher variance of the BBED could be mainly responsible for the improvements, as a higher variance potentially helps to generate better speech estimates. We also believe that too large values for the diffusion coefficient $g(t)$ may lead to numerical instability of the reverse process. We leave this discussion for future work.

\section{Conclusions} \label{sec:conclusion}
In this paper, we aimed to minimize the prior mismatch in score-based generative modeling for speech enhancement. To this end, we constructed the BBED SDE that is inspired by Brownian bridges. The BBED SDE yields a much smaller prior mismatch compared to the baseline OUVE SDE and has one hyperparameter less to tune. As a result, we consistently improve in all reported metrics over the OUVE SDE. Moreover, the BBED SDE achieves improvements of $0.13$ in PESQ and $0.26$ in POLQA even when only using half as many function evaluations as the OUVE SDE.

\bibliographystyle{IEEEtran}
\bibliography{ref}

\end{document}